\documentclass{epl}

\usepackage{amssymb}
\usepackage{epsfig}

\begin{document}

\title{Preon Trinity - A Schematic Model of Leptons, Quarks and
Heavy Vector Bosons}
\author{Jean-Jacques~Dugne\inst{1}
\and Sverker~Fredriksson\inst{2}
\and Johan~Hansson\inst{2}}
\institute{
  \inst{1} Laboratoire de Physique Corpusculaire,
Universit\'{e} Blaise Pascal, Clermont-Ferrand II,
FR-63177  Aubi\`{e}re, France\\
  \inst{2} Department of Physics, Lule\aa \ University of Technology,
SE-97187 Lule\aa , Sweden
}

\shorttitle{Preon Trinity}
\shortauthor{J.-J.~Dugne, S.~Fredriksson and J.~Hansson}

\maketitle

\begin{abstract}
Quarks, leptons and heavy vector bosons
are suggested to be composed of
stable spin-$1/2$ preons, existing
in three flavours, combined
according to simple rules. Straightforward
consequences of an $SU(3)$ preon-flavour
symmetry are the conservation of three
lepton numbers, oscillations and decays between
some neutrinos, and the mixing of the $d$ and $s$
quarks, as well as of the vector fields
$W^{0}$ and $B^{0}$. We find a relation
between the Cabibbo and Weinberg mixing angles,
and predict new (heavy) leptons, quarks and
vector bosons, some of which might be
observable at the Fermilab Tevatron and
the future CERN LHC. A heavy neutrino
might even be visible in existing data
from the CERN LEP facility.
\end{abstract}

\section{Introduction}

The phenomenological
success of the standard model
of quarks and leptons, and their
observed patterns, indicate that
there exist a more fundamental basis.
Here we present a simple preon model
where leptons, quarks and heavy vector bosons
are composite, and where many of the {\it ad hoc}
ingredients of the standard model are clear-cut
consequences of this inner structure. We limit the
present discussion to straightforward, qualitative
consequences of the model, and leave a more
quantitative analysis, requiring
extra assumptions, to future publications.

Although there is currently no {\it direct}
experimental evidence for (or against) preons,
there are quite a few phenomenological,
and logical, circumstances that
point at compositeness, some of which have
been known for long ~\cite{preons}:

\vskip0.1cm
\noindent $\bullet$ There are ``too many'' leptons
and quarks, but still a pattern among them.
Historically, such patterns
have led to ideas about compositeness, with the
quark model as a modern example,
and the periodic system as an older one.

\vskip0.1cm
\noindent $\bullet$ The least elegant
(mathematical) features of the standard model
are due to the weak gauge bosons being massive
(and unstable and of different charges).
They might be preon-antipreon states,
in analogy to the nuclear force being ``leaked'' by
quark-antiquark states. If so, there is no
fundamental weak force, nor a Higgs mechanism.
The photon-$Z^{0}$ mixing is equivalent to
the photon-$\rho^{0}$ mixing
in the vector-dominance model.

\vskip0.1cm
\noindent $\bullet$ An overlooked hint
is that most leptons and quarks are
{\it unstable}, which in our view
disqualifies them as fundamental
particles. Historically, decays of
``elementary'' particles have sooner
or later been interpreted as
processes among more fundamental objects.

\vskip0.1cm
\noindent $\bullet$ There are mixings
or oscillations, among some quarks, among the
vector fields $W^{0}$ and $B^{0}$ of the weak
interaction, and among some neutrinos.
This reminds about the chemical mixing of
isotopes and the quantum-mechanical one
of $K^{0}$ and $\bar{K}^{0}$ mesons,
both being due to compositeness.

\vskip0.1cm
\noindent $\bullet$ The conservation
(exact or approximate) of lepton numbers,
baryon number and weak isospin, are not yet
properly understood. When strong isospin and
strangeness were introduced for hadrons, on
similar grounds, the explanation later turned
out to be in terms of compositeness (quarks).

\vskip0.1cm
Early preon models focused on explaining
the lightest quarks ($u$ and $d$)
and leptons ($e$ and $\nu_e$)
with as few preons as possible. This means either
a minimal number of (two) different
preons, {\it e.g.}, the ``rishons''
~\cite{harari,shupe}, or a minimal number
of (two) preons inside a quark or lepton,
{\it e.g.}, the ``haplons'' ~\cite{fritzsch}.
Following these ideas we now present a
preon model for {\it all} leptons and quarks,
addressing some of the less understood
cornerstones of the standard model.
Examples are lepton-number conservation,
the Cabibbo-Kobayashi-Maskawa (CKM) mixings
~\cite{cabibbo,kobayashi} and the ``photonic''
behaviour of the $Z^{0}$ boson. We have been
much inspired by other models:

\vskip0.1cm
\noindent $\bullet$ The original quark model
~\cite{gellmann,zweig}, prescribing that
the hadrons known in the early 1960s can be
explained in terms of three quark flavours,
with an (approximate) $SU(3)$ symmetry.

\vskip0.1cm
\noindent $\bullet$ The rishon and haplon models, where
the lightest quarks and leptons contain, respectively,
three spin-$1/2$, or one spin-$1/2$ and one
spin-$0$ preon each.

\vskip0.1cm
\noindent $\bullet$ Diquark models (reviewed in
~\cite{anselmino}), often prescribing that
quarks pair up in total spin $0$.

\vskip0.1cm
\noindent $\bullet$ Supersymmetry, where spin-$1/2$
objects have spin-$0$ relatives, even if only
phenomenologically, {\it e.g.}, as a quark-diquark
``supersymmetry'' ~\cite{anselmino}.

\section{A trinity of preons}

\noindent A preon model for six leptons
and six quarks must have at least
five different preons in the sense of
the haplon model, {\it i.e.}, three
with spin $1/2$ and two with spin $0$,
or {\it vice versa}. A symmetric scheme
should have three of each, giving
nine leptons and nine quarks.
The spin-$1/2$ and spin-$0$ preons can be
arranged as partners of identical charges.
We suggest that the spin-$0$
ones are not fundamental, but tightly
bound ``dipreon'' pairs,
kept together by spin-dependent forces.
Calling the preons $\alpha$, $\beta$ and $\delta$,
with electric charges $+e/3$, $-2e/3$ and
$+e/3$ (by choice; there is an ambiguity
in the names preon and antipreon), we
get the simple, symmetric scheme in Table I.
Each preon has a ``supersymmetric'' partner,
which is the anti-dipreon formed by the other two
(anti)preons.

\begin{table}
\caption{The ``supersymmetric'' preon scheme.}
\begin{center}
\begin{tabular}{l|ccc}
charge & $+e/3$ & $-2e/3$ & $+e/3$\\
\hline
spin-$1/2$ preons & $\alpha $ & $\beta $ & $\delta $ \\
spin-$0$ (anti-)dipreons & $(\bar{\beta} \bar{\delta})$ &
$(\bar{\alpha} \bar{\delta})$ &
$(\bar{\alpha} \bar{\beta})$\\
\end{tabular}
\end{center}
\end{table}

The preons are conjectured to have the following properties:

\vskip0.1cm
\noindent $\bullet$ {\it Mass}:
One of the preons, say $\delta$,
must be much heavier than the other two,
since only six leptons, six quarks and three
heavy vector bosons have been seen so far.
Curiously enough, the two dipreons
with a $\delta$ must not be superheavy.
Rather, it seems as if $(\alpha \beta)$,
the partner of $\delta$, is somewhat
heavier than $(\alpha \delta)$ and $(\beta \delta)$,
which, in turn, might be a clue to the
underlying preon dynamics.

\vskip0.1cm
\noindent $\bullet$ {\it Spin and electric charge}:
These are just implemented on the preon
level, although an unambiguous
definition of spin (and mass)
requires a free particle. Maybe the
mass difference between
$\alpha$ and $\delta$ is related to
different, and strong, magnetic charges,
coupled to spin and electric charge.

\vskip0.1cm
\noindent $\bullet$ {\it QCD colour}:
Preons (and anti-dipreons) are ${\bf 3^{*}_{c}}$
representations of the $SU(3)_{c}$ of normal QCD,
so that leptons and vector bosons become colour
singlets and quarks colour triplets.

\vskip0.1cm
\noindent $\bullet$ {\it Stability and preon flavour}:
We assume that preons are stable.
Hence, a weak decay is merely a reshuffling of
preons into particles with a lower total mass.
The preon-flavour $SU(3)$ symmetry is
similar to the one of the original
quark model. Preons are $\bf 3_{f}$
and dipreons $\bf 3^{*}_{f}$.

\vskip0.1cm
\noindent $\bullet$ {\it ``Hypercolour''?}:
Some preon models rely on a QCD-like force
that binds preons into leptons and quarks,
although its group symmetry might be more
complicated than $SU(3)$. We will not restrict
the model to any specific preon dynamics at
this ``schematic'' stage.

\section{Leptons}

\noindent Leptons are preon-dipreon systems
in colour singlet
($\bf 3^{*}_{c} \otimes \bf 3_{c}
= \bf 1_{c} \oplus \bf 8_{c}$),
as in Table II, where each cell contains the
lepton built by a preon to the left and
a dipreon from the top line.
The following conclusions can be drawn:

\begin{table}
\caption{Leptons as three-preon states.}
\begin{center}
\begin{tabular}{c|ccc}
& $(\beta \delta)$
& $(\alpha \delta)$
& $(\alpha \beta)$ \\
\hline
$\alpha$
& $\nu_{e}$
& $\mu^{+}$ & $\nu_{\tau}$ \\
$\beta$ & $e^{-}$
& $\bar{\nu}_{\mu}$
& $\tau^{-}$ \\
$\delta$
& $\nu_{\kappa 1}$
& $\kappa^{+}$
& $\nu_{\kappa 2}$ \\
\end{tabular}
\end{center}
\end{table}

\vskip0.1cm
\noindent $\bullet$ The known leptons are reproduced,
although with an ambiguity in the scheme;
the ``$e$'' and ``$\tau$'' columns
can be interchanged.

\vskip0.1cm
\noindent $\bullet$ There are three new (superheavy)
leptons, all containing an ``isolated''
$\delta$. Two of these are neutrinos,
which must hence be heavier than half the
$Z^{0}$ mass (and unstable). There is no
other clear-cut mass-ordering, although the mass
seems to increase along all diagonals from upper
left to lower right.

\vskip0.1cm
\noindent $\bullet$ The three lepton numbers are
conserved in all known leptonic
processes (except neutrino oscillations - see below),
due to the conservation of preon flavour,
and assuming that dipreons do not split up.
Muon decay, $\mu^{-} \rightarrow
\nu_{\mu} + e^{-} + \bar{\nu}_{e}$,
is a typical example:
$\bar{\alpha}
(\bar{\alpha} \bar{\delta}) \rightarrow
\bar{\beta} (\bar{\alpha} \bar{\delta}) +
\beta (\beta \delta) +
\bar{\alpha} (\bar{\beta} \bar{\delta})$.
The dipreon goes into a $\nu_{\mu}$,
and the preon into a $W^{-} = \beta \bar{\alpha}$,
which then decays to $ e^{-} \bar{\nu}_{e}$.
A dipreon split-up would violate energy
conservation, and lepton number
is here equivalent to ``dipreon number''.
Hence there is no fourth lepton number for
$\kappa$, $\nu_{\kappa 1,2}$.
Rather, their decays {\it must} violate
also the normal lepton-number conservation,
{\it e.g.},
$\kappa^{+} \rightarrow \mu^{+} +
\nu_e + \bar{\nu}_{\tau}$.

\vskip0.1cm
\noindent $\bullet$ There is a fair chance
that at least one of $\nu_{\kappa 1}$ and
$\nu_{\kappa 2}$ is lighter than the top quark,
since one of them is in the same
``cell'' as the top (see Table III).
In particular, $\nu_{\kappa 2}$
can then be produced in $e^{+}e^{-}$ reactions even
at CERN LEP energies (up to $209$ GeV), and hence
might be visible in existing data. The production
channel would be $e^{+}e^{-} \rightarrow
\bar{\nu}_{e} + \nu_{\kappa 2}$ or
$\nu_{\mu} + \nu_{\kappa 2}$, exploiting
that $\nu_{e}$, $\bar{\nu}_{\mu}$ and $\nu_{\kappa 2}$
contain the same preons. The best chance to
discover $\nu_{\kappa 2}$ is through its
decay channels
$\nu_{\kappa 2} \rightarrow \mu^{+} + W^{-}$,
$e^{-} + W^{+}$, or $\nu_{e}/\bar{\nu}_{\mu} + \gamma$
(and similar for $\bar{\nu}_{\kappa 2}$).
The experimental signatures for $\nu_{\kappa 2}$
would therefore be a fast $\mu$ or $e$
plus two hadronic jets (probably from an
on-shell $W$), or a single high-energy $\gamma$.
It is noteworthy
that our model has no channels like
$e^{+}e^{-} \rightarrow
\bar{\ell}_{light} + \ell_{heavy}$,
with charged leptons.

\vskip0.1cm
\noindent $\bullet$ There is no ``family''
subscheme. The $SU(3)$ flavour structure
resembles the baryon octet
($\bf 3_{f} \otimes 3^{*}_{f} = 8_{f} \oplus 1_{f}$)
in the quark model. The singlet is a mixture
of the three neutrinos on
the main diagonal, while the other two
combinations are the lepton versions of
$\Sigma^{0}$ and $\Lambda^{0}$ in the
quark model.

\vskip0.1cm
\noindent $\bullet$ The scheme contains
$e^{-}$, $\tau^{-}$ and $\mu^{+}$
on equal footing.
It is noteworthy that the neutrino
(negative) helicity is carried
by negatively charged preons
($\alpha$ in $\nu_{e}$ and $\nu_{\tau}$;
$\bar{\beta}$ in $\nu_{\mu}$;
$\delta$ in $\nu_{\kappa 1}$ and
$\nu_{\kappa 2}$), and opposite for
antineutrinos.

\vskip0.1cm
\noindent $\bullet$ The neutrinos $\nu_e$,
$\bar{\nu}_{\mu}$ and $\nu_{\kappa 2}$
can mix into mass eigenstates, as they
have identical net preon flavours:
$\nu_e = \alpha (\beta \delta)$,
$\bar{\nu}_{\mu} = \beta (\alpha \delta)$
and $\nu_{\kappa 2} = \delta (\alpha \beta)$.
The ambigous shift of columns in Table II
would involve $\nu_{\tau}$
in the mixing instead of $\nu_{e}$.
A realistic set of mass eigenstates,
leading to oscillations, is
$|\nu_{1} \rangle = \cos \theta_{P}|\nu_{e/{\tau}} \rangle
- \sin \theta_{P}|\bar{\nu}_{\mu}\rangle$,
$|\nu_{2}\rangle = \sin \theta_{P}|\nu_{e/{\tau}}\rangle
+ \cos \theta_{P}|\bar{\nu}_{\mu}\rangle$
and $|\nu_{3}\rangle = |\nu_{\kappa 2}\rangle$.
There should also be decays, like
$\nu_{2} \rightarrow \nu_{1} + \gamma$ ~\cite{hansson},
or $\nu_{2} \rightarrow \nu_{1} + \phi$,
where $\phi$ is a light scalar,
such as a Goldstone boson ~\cite{barger},
or a scalar $(\beta \bar{\beta})$ or
$(\alpha \bar{\alpha})$.
The situation is hence theoretically complex, with
decays {\it and} oscillations, maybe involving
both wrong-helicity (= ``sterile''?) and normal
components (see also ~\cite{lindner}).
We will therefore study the neutrino
sector in some detail elsewhere,
taking into account the bulk of data from
solar and atmospheric neutrinos. Such an
analysis will require additional assumptions,
outside the definition of our preon model,
{\it e.g.}, about the mixing angle,
the oscillation and decay lengths (and channels)
and the whereabouts of wrong-helicity
neutrinos.

\vskip0.1cm
\noindent $\bullet$ Mixings of {\it charged}
leptons, and decays like
$\mu \rightarrow e+ \gamma$, cannot occur.

\section{Quarks}

\noindent Quarks are bound states of a
preon and an anti-dipreon in colour triplet
($\bf 3^{*}_{c} \otimes 3^{*}_{c}
= 3_{c} \oplus 6^{*}_{c}$), as shown in Table III.
The following list of quark properties takes up
several features that do not exist for leptons:

\begin{table}
\caption{Quarks as preon-anti-dipreon states.}
\begin{center}
\begin{tabular}{l|ccc}
& $(\bar{\beta} \bar{\delta})$
& $(\bar{\alpha} \bar{\delta})$
& $(\bar{\alpha} \bar{\beta})$  \\
\hline
$\alpha$
& $u$
& $s$ & $c$ \\
$\beta$
& $d$
& $X$ & $b$  \\
$\delta$
& $t/h$
& $g$ & $h/t$  \\
\end{tabular}
\end{center}
\end{table}

\vskip0.1cm
\noindent $\bullet$ All known quarks are reproduced
with their quantum numbers. There are
a few ambiguities in the scheme, {\it e.g.}, with
$t$ and $h$ interchanged.

\vskip0.1cm
\noindent $\bullet$ Quarks are not as fundamental
here as leptons. There is no obvious ``hypercolour''
that groups the quarks into overall singlets,
and preons in quarks might be just dynamically
bound, more like in a resonance than in
a ``real'' particle. A hint is that the lone
preon is in a colour-singlet configuration
with one of the antipreons inside the
anti-dipreon.

\vskip0.1cm
\noindent $\bullet$ There are three new quarks,
but only two of these ($g$ and $h$, for
``gross'' and ``heavy'') contain an isolated
$\delta$, while the third ($X$) does not.
On the other hand, the top quark is, conveniently,
among the superheavies, indicating that ``superheavy''
in our model means a few hundred GeV. That would
make the CERN LHC and the Fermilab Tevatron ideal for
discovering new leptons, quarks and heavy vector bosons,
with existing CERN LEP data as a thrilling 
possibility, as noted above.

\vskip0.1cm
\noindent $\bullet$ The quarks have no family
grouping either. The flavour-$SU(3)$
decomposition is into a sextet and an
antitriplet ($\bf 3_{f} \otimes \bf 3_{f}
= \bf 6_{f} \oplus \bf 3^{*}_{f}$).
The $\bf 3^{*}_{f}$ contains the three
quarks on the main diagonal, while the
$\bf 6_{f}$ contains the off-diagonal ones.
A sextet assignment for the {\it known}
quarks has been suggested by
Davidson \etal~\cite{davidson},
and speculated to be due to compositeness.

\vskip0.1cm
\noindent $\bullet$ The $X$ quark
with charge $-4e/3$ might pose a problem, but
also opens up for some new opportunities.
Neglecting the possibility that a ``light''
$X$ has escaped discovery, we can imagine two
reasons for its ``non-existence'':

a) The system
$X = \beta + (\bar{\alpha} \bar{\delta})$
might be {\it unbound}. It has the strongest internal
electric repulsion of all quarks and
leptons, due to the charge $-2e/3$ of
both preon and anti-dipreon. Quarks are
probably smaller than $0.001$ fm, and
the internal electric forces are substantial.

b) $X$ might be the discovered top quark.
This cannot be easily dismissed, since the
top charge is not known. The top quark was found
through its presumed decay channels
$t \rightarrow b + \ell^{+} + \nu$
and $t \rightarrow b + u + \bar{d}$,
where a few $b$ decays have been ``tagged''
by a charged muon. The situation is complex,
because an event contains the decay of a
$t \bar{t}$ pair into leptons and hadrons.
Comparable channels for $X$ would be
$\bar{X} \rightarrow \bar{b} + \ell^{+} + \nu$
and $\bar{X} \rightarrow \bar{b} + q_{1} + \bar{q}_{2}$,
so that the full $X \bar{X}$ decay would
give the same leptons and jets
as a $t \bar{t}$ decay, although with
another $b/\ell$ matching.
Hopefully, this issue will be settled soon
at the Tevatron ~\cite{baur}. Also Chang \etal
~\cite{chang} suggest that the top charge is $-4e/3$,
after a ``standard-model'' data analysis.

\vskip0.1cm
\noindent $\bullet$ Quark decays are more
complicated than lepton decays, not the least
because they can be studied in detail only in
hadronic decays. Assuming that dipreons do not
readily break up into two preons,
a quark decay can go through four types of processes:

(a) Preons annihilate or change place with
preons in another quark in the same hadron.
Some of these decays violate
lepton-number conservation, an ``invisible''
example being
$D^{0} \rightarrow \bar{\nu}_{\tau}+\nu_{e}$
through
$[\alpha(\bar{\alpha}\bar{\beta})][\bar{\alpha}(\beta \delta)]
\rightarrow \bar{\alpha}(\bar{\alpha}\bar{\beta})
+ \alpha(\beta \delta)$.

(b) The dipreon ends up in a lighter quark, {\it e.g.},
beta decay $d \rightarrow u + e^{-} + \bar{\nu}_{e}$
through $\beta(\bar{\beta}\bar{\delta}) \rightarrow
\alpha(\bar{\beta}\bar{\delta}) +
\beta(\beta \delta) +
\bar{\alpha}(\bar{\beta}\bar{\delta})$.
This is similar to a charged-lepton decay.

(c) The dipreon ends up in a lighter {\it lepton}.
This does not conserve lepton numbers,
and involves a dipreon exchange
instead of a preon-antipreon exchange.
An example is $b \rightarrow
\bar{\nu}_{\tau} + e^{-} + u$ through
$\beta(\bar{\alpha}\bar{\beta}) \rightarrow
\bar{\alpha}(\bar{\alpha}\bar{\beta})+
\beta(\beta \delta) + \alpha(\bar{\beta} \bar{\delta})$.
Such decays are associated with the smallest
CKM mixings, reflecting that dipreon exchange is
suppressed. A problem here is
that a $c$ quark {\it must} decay to $s$, $u$ or
$d$ through such channels, {\it e.g.},
$c \rightarrow \bar{\nu}_{\tau} + s + \mu^{+}$.

(d) A preon annihilates an antipreon
in the same quark, resulting in another
preon-antipreon pair inside a lighter quark.
Such transitions can occur between quarks with
identical net preon flavours, such as $d$ and $s$.
Hence $s \leftrightarrow d$ goes through
$\alpha \bar{\alpha} \leftrightarrow \beta \bar{\beta}$.
This, in turn, leads to a quantum-mechanical
(Cabibbo) mixing of the two quarks
into two mass eigenstates. Such a mixing
is important also for the $Z^{0}$ (see below).
In terms of wave functions and the
notions of the Cabibbo theory, the mass
eigenstates are
\begin{equation}
|d'\rangle =  \cos \theta_{C}|\beta(\bar{\beta}\bar{\delta})\rangle
+ \sin \theta_{C}|\alpha(\bar{\alpha}\bar{\delta})\rangle
\end{equation}
\begin{equation}
|s'\rangle = \cos \theta_{C} |\alpha(\bar{\alpha}\bar{\delta})\rangle
- \sin \theta_{C} |\beta(\bar{\beta}\bar{\delta})\rangle.
\end{equation}

In the ($d'$) ground-state $\beta \bar{\beta}$
dominates over $\alpha\bar{\alpha}$ by
roughly a factor four, which might
be due to a stronger electric attraction in
the $\beta \bar{\beta}$ system.
In the quark model $u\bar{u}$ and $d\bar{d}$
have the same weight in the lightest
mesons ($\pi^{0}$ and $\rho^{0}$), because a meson is much
more extended than a quark, and hence electric forces
are less important there.

\vskip0.1cm
\noindent $\bullet$ As within some other preon
models the normal hydrogen atom here contains as many
antipreons as preons. The proton has the preon flavour of
$e^{+} + 2(\alpha \bar{\delta})$, and {\it decays},
{\it e.g.}, through $p \rightarrow
e^{+} + 2 \bar{\nu}_{e} + 2 \nu_{\tau}$.
This requires a complicated rearrangement
of three preons into three antipreons and
three neutrinos.

\section{Heavy vector bosons}

\noindent There are nine preon-antipreon states in the
shape of vector bosons, as in Table IV.
The following observations can be made:

\begin{table}
\caption{Heavy vector bosons as preon-antipreon states.}
\begin{center}
\begin{tabular}{l|ccc}
& $\bar{\alpha}$
& $\bar{\beta}$
& $\bar{\delta}$  \\
\hline
$\alpha$
& $Z^{0},Z'$ & $W^{+}$ & $Z^{*}$ \\
$\beta$
& $W^{-}$ & $Z',Z^{0}$ & $W'^{-}$  \\
$\delta$
& $\bar{Z}^{*}$ & $W'^{+}$ & $Z'',Z'$  \\
\end{tabular}
\end{center}
\end{table}

\vskip0.1cm
\noindent $\bullet$ The scheme is similar to the
vector meson octet in the quark model
($\rho,\omega,\phi,K^{*}$). Both carry a
``leakage force''; the weak and nuclear ones,
and both $Z^{0}$ and $\rho^{0}$ ``mix'' with
the photon through their constituents.
There is hence a ``vector meson dominance''
of the photon also in the preon sector.

\vskip0.1cm
\noindent $\bullet$ $W$ decays are split-ups into
other preon states, like $W^{-} \rightarrow e^{-} +
\bar{\nu}_{e}$ through $\beta \bar{\alpha} \rightarrow
\beta(\beta \delta) + \bar{\alpha}(\bar{\beta}\bar{\delta})$.
Such channels dominate also $Z^{0}$ decay, but
there are, in addition, annihilation
channels inside $Z^{0}$, {\it e.g.},
$\beta \bar{\beta} \rightarrow \gamma^{*}
\rightarrow \ell \bar{\ell}$.

\vskip0.1cm
\noindent $\bullet$ Weak isospin and the Weinberg mixing appear in
the model for the same reason as the Cabibbo and neutrino
mixings, {\it i.e.}, as consequences of preon-flavour
$SU(3)$, which mixes states with identical net preon
flavours into mass eigenstates. The weak $SU(2)$
isotriplet wave function is supposed to be
$|W^{0} \rangle = (|\alpha \bar{\alpha} \rangle
- |\beta \bar{\beta} \rangle)/\sqrt{2}$,
while the isosinglet is
$|B^{0} \rangle = (|\alpha \bar{\alpha} \rangle
+ |\beta \bar{\beta} \rangle)/\sqrt{2}$.
In terms of the Weinberg angle,

\begin{equation}
|Z^{0} \rangle = -\frac{1}{\sqrt{2}}(\cos\theta_{W} + \sin\theta_{W})
|\beta \bar{\beta}\rangle
+ \frac{1}{\sqrt{2}}(\cos\theta_{W} - \sin\theta_{W})
|\alpha \bar{\alpha}\rangle.
\end{equation}

Suppose that the
$\beta \bar{\beta} \leftrightarrow \alpha \bar{\alpha}$
admixture depends only on the $\beta$ and
$\alpha$ electric charges. Then we expect the
admixture (up to a sign set by convention
in the definitions of the Cabibbo and Weinberg angles)
to be the same in the ground-states $|d'\rangle$ and
$|Z^{0}\rangle$:

\begin{equation}
\cos\theta_{W} - \sin\theta_{W} = \sqrt{2}\sin\theta_{C}.
\end{equation}

With $\sin^{2}\theta_{W} = 0.23117 \pm 0.00016$ and
$\sin\theta_{C} = 0.2225 \pm 0.0035$
~\cite{rpp} the $lhs = 0.396 \pm 0.001$
and the $rhs = 0.315 \pm 0.005$. This is a
fair agreement considering the rough assumptions.

\vskip0.1cm
\noindent $\bullet$ The nearest orthogonal partner
of the $Z^{0}$ is a heavier $Z'$.
One cannot profit from the
analogy with the vector mesons $\rho^{0}$ and
$\omega$, and predict that $Z'$ will be near
$Z^{0}$ in mass, since the unequal
$\beta \bar{\beta}\leftrightarrow \alpha \bar{\alpha}$
admixture differs from the quark situation in
vector mesons. In addition, $Z'$ might
have a superheavy $\delta \bar{\delta}$
component, in analogy with scalar mesons,
where the $s\bar{s}$ component makes the
$\eta$ much heavier than the $\pi^{0}$.

\vskip0.1cm
\noindent $\bullet$ Scalar counterparts to the
vector bosons are expected in many preon models
~\cite{preons}. It is not known
if these are even heavier than the vector bosons,
or have an extra weak coupling to other particles.

\section{Conclusions and outlook}

\noindent Our model, as defined by the preon
schemes of Tables I-IV, provides a qualitative understanding of
several phenomena that are not normally addressed
by preon models, nor explained within the standard model.
It can hence serve as a basis for a deeper analysis
of the many disjunct ingredients of the latter. In addition,
the model predicts several new leptons and quarks,
which might be discovered in the near future,
maybe even in existing data from the (closed)
CERN LEP facility.

However, many problems remain to be solved:
The model still lacks a dynamics, so that masses,
reaction rates, etc.,
cannot yet be reproduced or predicted.
In this respect we are worse off
than the original quark model, which had at least
a phenomenological mass formula for baryons.
Also, it is not yet clear if {\it all}
phenomenologically successful aspects of
the electroweak theory can be explained by preons.
More work is obviously needed before a quantitative
theory may emerge. This will include making
extra assumptions, in addition to the mere
``observations'' made here, and analysing
the data in the light of these.
We will continue our efforts
with a deeper analysis of the neutrino sector,
in order to find oscillation and
decay patterns. A thorough analysis of existing
CERN LEP data would also be desirable,
in search of decay products of a heavy neutrino,
along the lines predicted by our model.

\acknowledgments
We acknowledge illuminating discussions
and correspondences with D. Chang, A. Davidson,
H. Fritzsch, D. Lichtenberg, E. Ma, B. McKellar,
E. Predazzi, R. Volkas and N. Tracas, as well as
valuable experimental information from G. Bellettini,
R. Partridge and G. VanDalen. This project has been
supported by the European Commission under contract
CHRX-CT94-0450, within the network
``The Fundamental Structure of Matter".

\end{document}